\documentclass{article}
\usepackage{spconf,amsmath,graphicx,cite,subfigure,multirow,booktabs,tabu,bm,amssymb}
\usepackage[bookmarks=true]{hyperref}

\title{Cross-speaker Emotion Transfer Based on Speaker Condition Layer Normalization and Semi-Supervised Training in Text-To-Speech}
\name{Pengfei Wu, Junjie Pan, Chenchang Xu, Junhui Zhang, Lin Wu, Xiang Yin, Zejun Ma}
\address{AI Lab, ByteDance \\
    \tt\small{\{wupengfei.ganyue, panjunjie.jeff, xuchenchang\}@bytedance.com}
}

\begin{document}
    %
    \maketitle
    \begin{abstract}
        In expressive speech synthesis, there are high requirements for emotion interpretation. However, it is time-consuming to acquire emotional audio corpus for arbitrary speakers due to their deduction ability. In response to this problem, this paper proposes a cross-speaker emotion transfer method that can realize the transfer of emotions from \emph{source speaker} to \emph{target speaker}. A set of emotion tokens is firstly defined to represent various categories of emotions. They are trained to be highly correlated with corresponding emotions for controllable synthesis by cross-entropy loss and semi-supervised training strategy. Meanwhile, to eliminate the down-gradation to the timbre similarity from cross-speaker emotion transfer, speaker condition layer normalization is implemented to model speaker characteristics. 
        Experimental results show that the proposed method outperforms the multi-reference based baseline in terms of timbre similarity, stability and emotion perceive evaluations. 
        
    \end{abstract}
    
    \begin{figure*}[h]
        \centering
        \subfigure[Proposed Model]{\label{fig:main_structure}
            \includegraphics[width=0.58\linewidth]{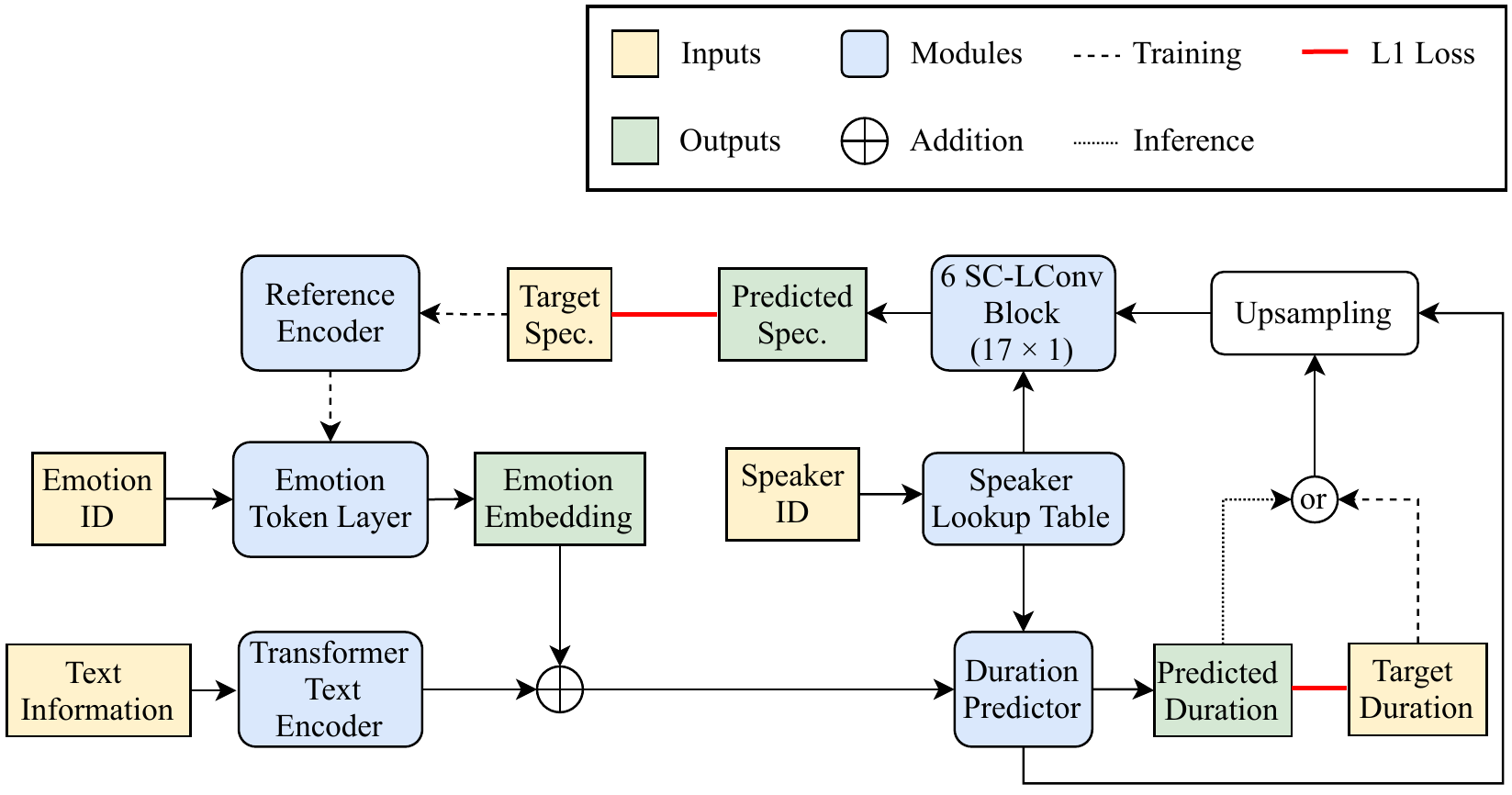}}
        \subfigure[Duration Predictor]{\label{fig:Duration_Predictor}
            \includegraphics[width=0.19\linewidth]{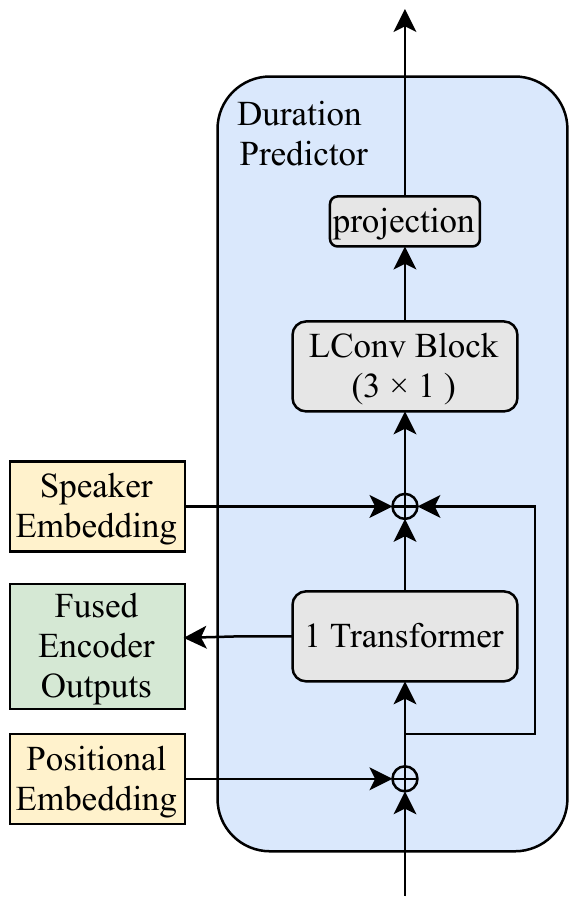}}
        \subfigure[SCLN-LConv Block]{\label{fig:SC_LConv_Block}
            \includegraphics[width=0.19\linewidth]{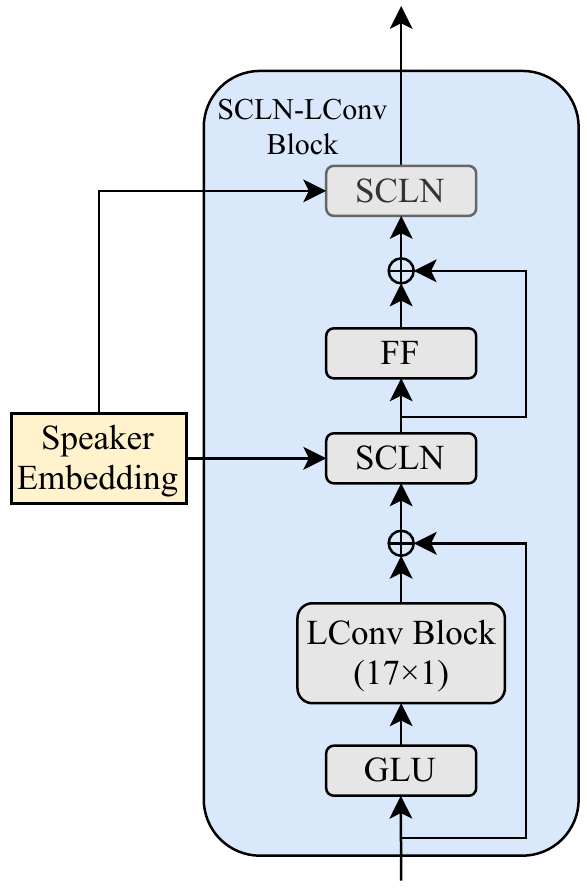}}
        \caption{The architecture of (a) proposed model, (b) duration predictor, (c) speaker condition layer normalization lightweight convolution block.}
    \end{figure*}
    \section{Introduction}
    \label{sec:intro}
    
    Text-to-speech (TTS) aims to produce natural-sounding speech. 
    In recent years, various deep learning based TTS acoustic methods\cite{YF2014SSbiLSTM,SoteloMKSKCB17,Wang17tacotron,ArikCCDGKLMNRSS17,Shen18tacotron2} and vocoder methods\cite{OordDZSVGKSK16,MehriKGKJSCB17,KalchbrennerESN18} are proposed, to generate high quality speech.
    However, human speech contains a lot more super-segmental information beyond texts, such as prosody and emotion, which are essential to further improve the naturalness of the synthesized speech.
    
    Several style modeling methods\cite{AnLD17,abs-1711-05447,Skerry18tacotron,Wang18GSTs,AkuzawaIM18,ZhangPHL19,abs-1904-02373,WuLLJWD19,multi-reference,HabibMSBSSKB20,abs-2109-06733} are proposed to model these non-textual information. Part of them highly depend on data with extra annotations\cite{AnLD17,abs-1711-05447}, which are complicated and lack generality and consistency, making it impractical in commercial production. 
    On the other hand, some propose unsupervised methods\cite{Skerry18tacotron,Wang18GSTs,AkuzawaIM18,ZhangPHL19} with encoder-decoder architectures, where utterance level representations are extracted by a style or prosody encoder, to boost the expressiveness of synthesized speech. However, the learned representations usually lack interpretability and controllability. Semi-supervised methods\cite{WuLLJWD19,HabibMSBSSKB20} are therefore proposed to improve the interpretability of the learned representations by providing partial supervision. 
    Wu et al. \cite{WuLLJWD19} propose an emotion control method called Semi-GST with 5\% supervision data which heuristically turns style token weights into one-hot vectors by introducing a semi-supervised cross-entropy loss. 
    In this manner, a specific physical interpretation (a single emotion) can be assigned to one style token. 
    Recently, multi-reference methods \cite{abs-1904-02373,multi-reference} are proposed for better performance and interpretation. They tried to learn independent speaker and style representations through multiple reference encoders. 
    By introducing strategies such as inter-cross training, paired-unpaired triplets and adversarial cycle consistency, those methods achieve the purpose of learning independent speaker and style representations.
    
    This paper aims to transfer emotions from \emph{source speaker} with multi-emotional speech corpus to the \emph{target speaker} without emotional annotations. Two main issues exist in this kind of emotion transfer task, that in the \emph{target speaker} synthesis, the emotion perception and pronunciation stability should be guaranteed, and the timbre similarity should be kept. To resolve that, we propose a parallel Tacotron\cite{parallel-taco}  based model, with the variational residual encoder replaced by a global style tokens(GSTs) module because we are aiming at controllable cross-speaker emotion transfer.
    The contributions of this paper include the following. Firstly, we propose to use GSTs and semi-supervised training strategy for controllable cross-speaker emotion transfer. 
    Secondly, we introduce speaker condition layer normalization (SCLN) into cross-speaker emotion transfer task. 
    
    We notice that recent work by Li et al. \cite{abs-2109-06733} which achieves cross-speaker emotion transfer based on Tacotron 2. The proposed work differs from Li's as follows. Firstly, their back-bone is an autoregressive model, while our back-bone is non-autoregressive. We believe that non-autoregressive models perform better in feature decoupling because they do not directly take the previous frame as input when predicting the next one, resulting in less feature leakage. Secondly, they get speaker-independent emotion embeddings by explicitly constraining speaker and emotion embeddings, while we learn speaker embeddings by SCLN blocks, and emotion embeddings by emotion tokens and semi-supervised strategy.

    \section{Proposed Method}
    \label{sec:proposed_method}
    The architecture of the proposed cross-speaker emotion transfer model is illustrated in Fig.~\ref{fig:main_structure}. 
    It mainly consists of a Transformer-based text encoder, a Transformer-based duration predictor, an upsampling block, a spectrogram decoder stacks, a reference encoder, and an emotion token layer. 
    \vspace{-0.2cm}
    \subsection{Duration Predictor}
    The duration predictor takes the encoder outputs added by the emotion embedding as input, and outputs the predicted phoneme durations. We extract the ground-truth phoneme durations, by an external hidden Markov model(HMM)-based aligner, to train
    the duration predictor. As shown in Fig. \ref{fig:Duration_Predictor}, it consists of a Transformer block, a 3×1 LConv block, and a projection layer.  A sinusoidal positional embedding is added to the inputs and then fed into the Transformer block. The outputs of the Transformer block are used for upsampling and added to speaker embedding as the input of the LConv block. At last, the speaker and emotion-related phoneme durations are predicted by the LConv block and the projection layer. 

    \vspace{-0.2cm}
    \subsection{SCLN-LConv Block}
    The spectrogram decoder consists of six SCLN-LConv blocks, and the architecture of SCLN-LConv block is shown in Fig. \ref{fig:SC_LConv_Block}. We insert the SCLN module after the LConv block and FF layer in each block. The SCLN module is a conditional layer normalization\cite{adaspeech} which takes the speaker embedding as inputs and predicts the scale and bias parameters of layer normalization. 
    \vspace{-0.2cm}
    \subsection{Emotion Token Layer and Semi-Supervised Training}
    We model emotion properties by introducing utterance-level emotion embeddings, which are extracted as follows. Firstly, the target mel-spectrogram is fed into a reference encoder, which encodes the reference mel-spectrogram into a fixed-length vector called reference embedding. Thereafter, the reference embedding is used as query to calculate a set of weights with pre-defined emotion tokens using a single-head attention module. Finally, the emotion embedding is generated by the weighted sum of the emotion tokens.   
    
    Similar to Wu et al.\cite{WuLLJWD19}, we add an emotion classifier loss between token weights and one-hot emotion ID to ensure that the trained tokens have a one-to-one correspondence with emotions at the training stage. In this way, the emotion embedding can be generated by multiplying one-hot emotion ID and emotion tokens during the inference stage. 
    Since this paper focuses on cross-speaker emotion transfer based on disjoint databases, it is worth considering how to deal with the situation where the \emph{target speaker} has no emotion annotations. 
    Instead of regarding all emotions of \emph{target speaker's} speech as \emph{neutral}\cite{multi-reference},  we treat it as a semi-supervised learning problem. The emotion classifier loss of \emph{target speaker} is not calculate and the model will softly determine what emotions each speech contains.
    
    Overall, the training objective of the proposed method are shown in Eq.~\ref{eq1} $ \sim $ Eq.~\ref{eq4},
     \begin{equation}\label{eq1}
        \mathcal{L}_{ec} =  - \sum_{i \in  s_{s}}\textbf{e}_{i}log(\hat{\textbf{e}}_{i}) 
    \end{equation}
    \begin{equation}\label{eq4}
        \mathcal{L} = \sum_{i \in K} \mathcal{L}_{reco}^{i} + \alpha \mathcal{L}_{ec} + \beta \mathcal{L}_{dur}
    \end{equation}
    where $\mathcal{L}_{reco}^{i}$ is the reconstruction loss of the $i$-th decoder stack, $ K $ is the number of decoder stacks, $\mathcal{L}_{dur}$ is the duration loss, $\mathcal{L}_{ec}$ is the emotion classifier loss, $ s_{s} $ denotes to \emph{source speaker}, $ \alpha $ and $ \beta $ is the loss weight of emotion classifier loss and duration loss respectively.
    
    \section{Experiments}
    \label{sec:experiments}
    \subsection{Experimental Setup}
    \label{sec:exp_setup}
    
    Two internal Chinese speech databases from two male speakers are utilized in our experiments. One is a multi-emotion speech database with 7-emotion annotations (refer as \emph{source speaker}), and the other is an audio-book database (refer as \emph{target speaker}). 
    The \emph{source speaker} database contains 7-emotion annotations (800 utterances in each) and is 8.32 hours in total. 
    The \emph{target speaker} database contains 6778 utterances, and the total duration is 8.61 hours.
    80-dimensional mel-frequency spectrograms are extracted with 10ms frame shift and 50ms frame length. We split 50 utterances in each corpus for the test.
    To evaluate the performance of the proposed method, the following models are constructed for further comparisons.
    \begin{itemize}
        \item \textbf{baseline}: A parallel tacotron-based multi-reference emotion transfer model using the paired-unpaired training strategy and adversarial cycle consistency scheme proposed by Whitehill et al.\cite{multi-reference}. The emotion embeddings and the speaker embeddings generated by reference encoders are concatenated with encoder outputs.
        
        \item \textbf{proposed}: Proposed model describe in Sec. \ref{sec:proposed_method}.
        
        \item \textbf{M1}: An ablation model which removes the SCLN module in decoder LConv blocks and the speaker embeddings are added to encoder outputs.
        
        \item \textbf{M2}: An ablation model which removes emotion classifier loss in the training stage, multi-head attention is utilized in this model.

    \end{itemize}

    In the \textbf{proposed} and \textbf{M1}, seven 256-dimensial emotion tokens are pre-defined, and $ \alpha $ and $\beta $ in Eq.~\ref{eq4} are both  set to 0.1. As for \textbf{M2}, tokens and heads are set to 10 and 4, respectively. Speaker IDs are mapped into 64-dimensional vectors with a speaker lookup table. The \textbf{baseline} is implemented following the setup of \cite{multi-reference} except the change of back-bone. 
    All models are trained with 32 batch size for 200k steps. WaveRNN\cite{KalchbrennerESN18}  is used in our experiments as the vocoder.
    \subsection{Results and Analysis}
    \label{sub:results_analysis}
    In this paper, we conduct three types of subjective evaluations to compare the cross-speaker emotion transfer performance of different models. All the samples are synthesized using unseen texts.
    
    \textbf{Timbre similarity}: Participants are given synthesized speech of \emph{target speaker} and two original recordings from \emph{source speaker} and \emph{target speaker} respectively. They are asked to give a 1$ \sim $5 score with 0.5 interval for the timbre similarity between each synthesized speech and recordings. Higher score means higher similarities with the \emph{target speaker}'s timbre. In our experiments, 70 utterances (10 for each emotion) are synthesized for each model, and 15 participants conduct this evaluation. Furthermore, utterances in the same group are shuffled and the model information is invisible to participants.

    \textbf{Stability comparison feedback}: Given the  synthesized speech from different models, participants are asked to give feedback if there are stability problems such as missing words, speaking rate problem, blurred speech, and pronunciation defect. The results are the percentages of synthesized speech with stability problems in sentence level. In this evaluation, 300 utterances (60 for \emph{neutral} and 40 for the rest of emotions) are synthesized from each model and one linguistic expert conducts this evaluation. The model information is invisible to participant.
     
    \textbf{Emotion perceive preference}: It is carried out in the form of ABX test. Speech are synthesized from both models with the same emotion labels and texts. Participants are asked to determine which utterance is perceived closer to the description of the emotion label. The two utterances with the same text and emotion label are scored in parallel with the random order, and model information is invisible to participants. The set of synthesized speech used in this evaluation are the same as in the timbre similarity evaluation.

    \subsubsection{Comparison with baseline}
    In order to compare with the baseline, three subjective evaluations aforementioned and an objective timbre similarity evaluation are conducted. The results are shown in Table \ref{table:timbre_similarity} $ \sim $ Table \ref{table:emotion_perveive_preference} and Fig. 2.
    
\begin{figure}[t]
        \subfigure[SV embeddings of the baseline.]{\label{fig:sv_baseline}
            \includegraphics[width=0.48\linewidth]{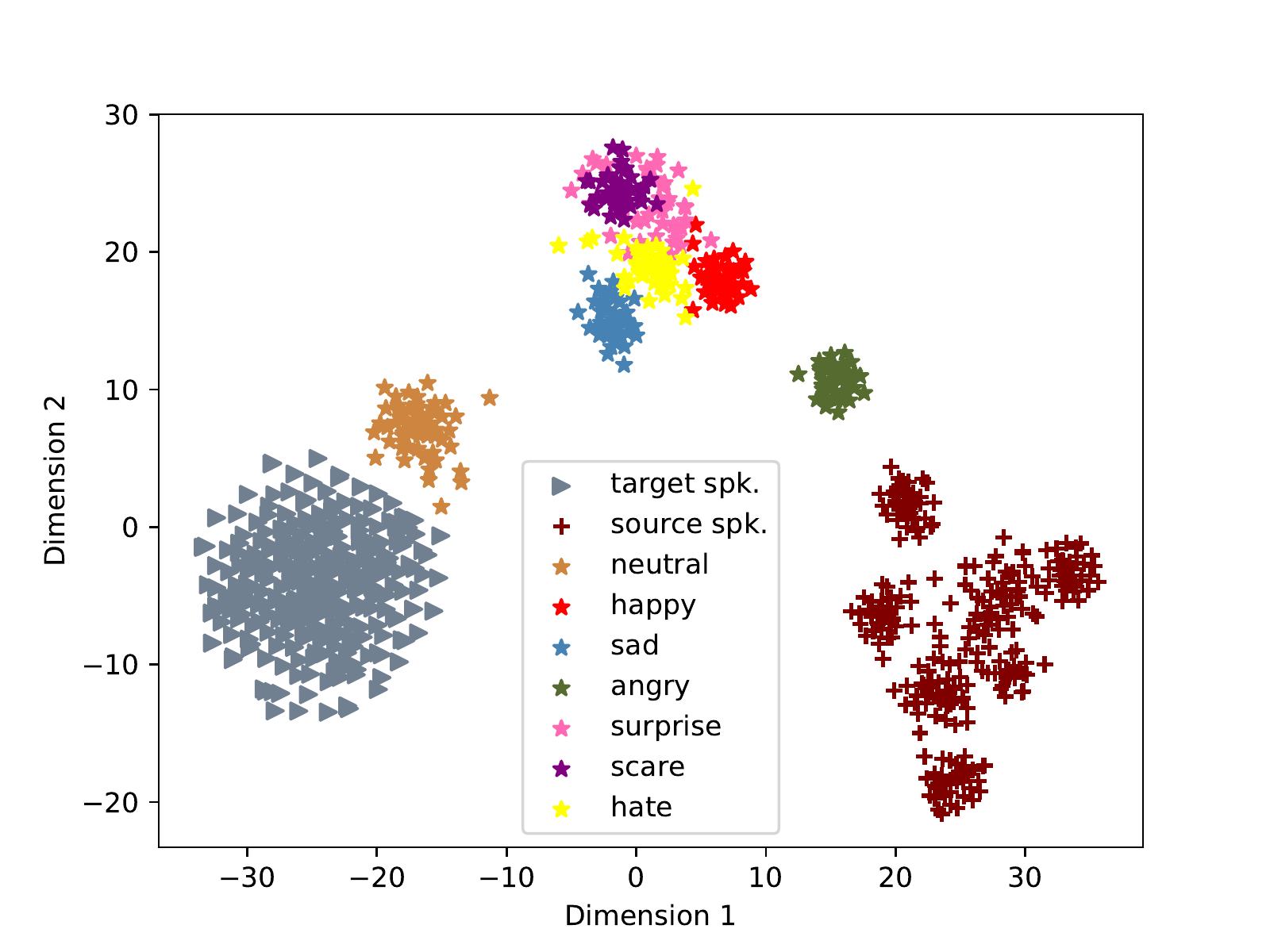}}
        \subfigure[SV embeddings  of the proposed.]{\label{fig:sv_proposed}
            \includegraphics[width=0.48\linewidth]{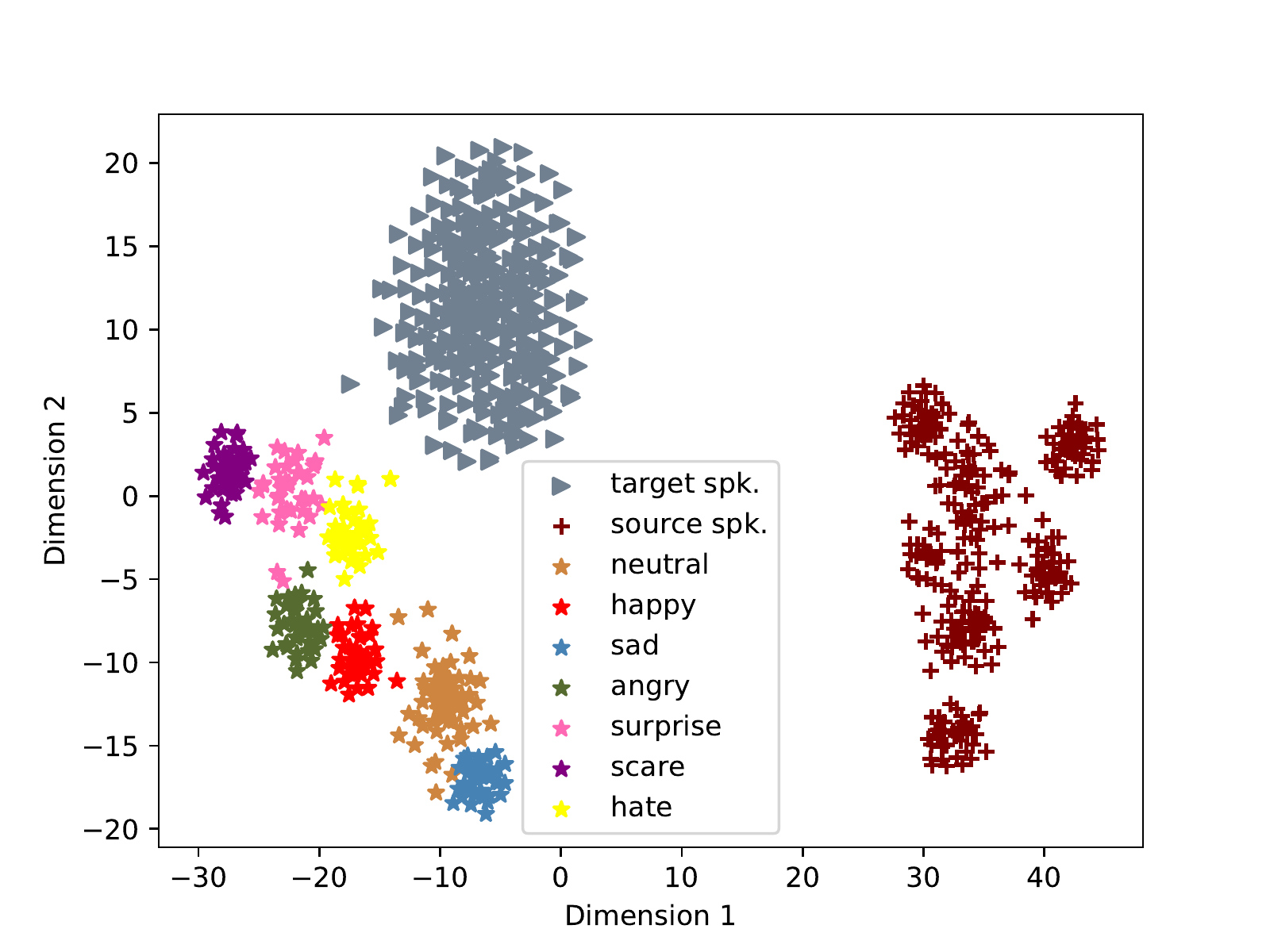}}
        \caption{The SV embeddings of different models, each point corresponds to one SV embedding. '$\blacktriangleright$' and '$ + $' denote \emph{target speaker's} and \emph{source speaker's} SV embedding points respectively, '$ \star$' denotes SV embeddings of synthesized speech and colors represent different emotions.}
    \vspace{-0.2cm}
    \end{figure}

    Table \ref{table:timbre_similarity} shows that the proposed method outperforms the baseline in terms of subjective timbre similarity. Especially, the baseline gets extremely lower scores in \emph{sad}, \emph{angry}, \emph{surprise} and \emph{scare}. 
    To objectively compare the timbre similarity of baseline and the proposed method, we randomly select 300 recordings from \emph{target speaker} and \emph{source speaker}, and then extract the 1024-dimensional utterance-level speaker verification (SV) embeddings of both recordings and synthesized speech using a pre-trained SV model.  The SV embeddings are reduced to 2-dimensional vectors using t-SNE\cite{van2008visualizing} and are plotted in Fig. \ref{fig:sv_baseline} and Fig. \ref{fig:sv_proposed}.  
    As shown in these figures, the synthesized clusters of the proposed method are closer to the \emph{target speaker} than baseline, indicating a better objective performance in timbre similarity. 
    Fig. \ref{fig:sv_baseline} shows that the \emph{neutral} cluster is closer to the \emph{target speaker} cluster, and the \emph{angry} cluster is closer to the \emph{source speaker} cluster. It means that the baseline performs better in \emph{neutral} and worse in \emph{angry}, which highly matched the subjective results in Table \ref{table:timbre_similarity}. 
    The objective results again prove that the proposed method performs better than the baseline in the timbre similarity.
    Moreover, it observes that there are several relatively separated clusters in the recordings of the \emph{source speaker}, proving that even for the same speaker, his/her timbre changes slightly in different emotions. This observation gives us an inspiration that small timbre changes in different emotions should be reasonable in the emotion transfer task.
    
    \begin{table}[t]
        \begin{center}
        \caption{Subjective timbre similarity evaluation results of different model, with confidence intervals of 95\%. The higher value means the better timbre similarity and the bold indicates the best performance in all the models. } 
        \resizebox{\columnwidth}{!}{
            \begin{tabular}{lcccc}
               \toprule
                \textbf{emotions} & \textbf{baseline}        & \textbf{M1}            & \textbf{M2}     & \textbf{proposed} \\ 
                \midrule
            neutral  & \textbf{4.05} $\pm$ 0.14 & 3.19 $\pm$ 0.13 & 3.91 $\pm$ 0.15 & 3.50 $\pm$ 0.13      \\ 
            happy    & 3.37 $\pm$ 0.13 & \textbf{4.03} $\pm$ 0.05 & 3.77 $\pm$ 0.13 & 4.00 $\pm$ 0.07       \\ 
            sad      & 3.00 $\pm$ 0.13 & 2.89 $\pm$ 0.09 & \textbf{3.96} $\pm$ 0.16 & 3.36 $\pm$ 0.09  \\ 
            angry    & 2.96 $\pm$ 0.06 & 3.21 $\pm$ 0.09 & 3.33 $\pm$ 0.11 & \textbf{3.67} $\pm$ 0.10    \\ 
            surprise & 3.16 $\pm$ 0.19 & \textbf{3.91} $\pm$ 0.09 & 3.85 $\pm$ 0.09 & 3.76 $\pm$ 0.15   \\ 
            scare    & 3.11 $\pm$ 0.11 & 3.35 $\pm$ 0.07 & \textbf{3.52} $\pm$ 0.09 & 3.48 $\pm$ 0.12  \\ 
            hate     & 3.33 $\pm$ 0.22 & \textbf{4.02} $\pm$ 0.11 & 4.01 $\pm$ 0.12 & 3.85 $\pm$ 0.13    \\ 
                \midrule
            average  & 3.28 $\pm$ 0.10 & 3.51 $\pm$ 0.11 & \textbf{3.76} $\pm$ 0.07 & 3.66 $\pm$ 0.07  \\ 
            \bottomrule
            \end{tabular}
            }
            \label{table:timbre_similarity}
        \end{center}
    \end{table}
    
    \begin{table}[t]
        \begin{center}
        \caption{Stability comparison feedback error rates of different models (\%), because one utterance may contains multiple stability problems, the \emph{total} may is not equal to the sum of the first 4 rows. The lower value means the better stability performance and the bold indicates the best performance in all the models.} 
            \begin{tabular}{ccccc}
            \toprule
            \textbf{stability problems}   & \textbf{baseline} & \textbf{M2} & \textbf{proposed} \\ 
             \midrule
            missing words         & 0.67             & \textbf{0.33}            & 0.67    \\ 
            speaking rate         & 6.67             & 6.00            & \textbf{4.67}       \\ 
            blurred speech        & 25.67            & 20.67           & \textbf{10.00}      \\ 
            pronunciation defect         & 5.67             & \textbf{2.67}            & 6.00       \\ 
            \midrule
            total                 & 32.67            & 27.30           & \textbf{18.33}      \\ 
            \bottomrule
            \end{tabular}
            \label{table:stability_comparison}
        \end{center}
    \end{table}

    \begin{table}[t]
        \begin{center}
        \caption{Average preference scores (\%) of the emotion perceive evluations, where N/P stands for “no preference”, and $\bm{p}$ denotes the $\bm{p}$-value of a $\bm{t}$-test  between two models. The higher value means stronger preference.} 
            \begin{tabular}{ccccc}
            \toprule
            \textbf{baseline} & \textbf{M2} & \textbf{proposed} & \textbf{N/P}   & \textbf{$\bm{p}$}         \\ 
            \midrule
            40.57    & -                & \textbf{53.24}    & 6.19 & \textless{}0.01 \\ 
            -        & 46.10            & \textbf{46.57}    & 7.33 & 0.87          \\ 
            \bottomrule
            \end{tabular}
            \label{table:emotion_perveive_preference}
        \end{center}
    \vspace{-0.2cm}
    \end{table}

    Table \ref{table:stability_comparison} shows that the stability performance of the proposed method is much better than the baseline, especially in percentage of blurred speech. 
    Table \ref{table:emotion_perveive_preference} shows that the proposed method significantly outperforms the baseline in emotion perceive preference test( $ p \textless 0.01 $). 
    In fact, the synthesized speech of the baseline can express the corresponding emotions correctly~\footnote{Demos can be found at: \url{https://acmlxg.github.io/icassp2022/}}, but that of our proposed method is even stronger and more accurate.
    
    \subsubsection{Ablation Evaluations}
    We conduct two ablation evaluations to demonstrate the effect of the SCLN-LConv blocks and semi-supervised strategy. 
    Firstly, we evaluate the performance of the SCLN-LConv blocks by comparing the proposed method with M1 in terms of timbre similarity. As shown in Table \ref{table:timbre_similarity}, the proposed method has an overall timbre similarity score improvement of 0.15 compared to M1.
    Especially, the proposed method performs much better than M1 in \emph{sad} and \emph{angry}.
    Then, we evaluate the effect of semi-supervised strategy by comparing the proposed method with M2. We randomly select one sample from the \emph{source speaker's} test set as reference for each emotion, and synthesize the evaluated samples. As shown in Table \ref{table:timbre_similarity} $ \sim $ Table \ref{table:emotion_perveive_preference}, M2 outperforms slightly than the proposed method in terms of timbre similarity, without  significant difference ( $ p = 0.87 $) in terms of emotion perception. However, in terms of stability, the total error rate of M2 is 9\% higher than the proposed method in absolute value.  Considering that stability is more essential for an online {TTS} system in large-scale commercial production, the proposed method is chosen as the final configuration.

    \section{Conclusion}
    \label{sec:conclusion}
    
    In this paper, we propose a cross-speaker emotion transfer method based on semi-supervised training and SCLN. An semi-supervised emotion classifier loss is introduced for the emotion interpolation in style tokens, and speaker condition layer normalization module is implemented to reserve speaker characteristics during cross-speaker emotion transfer. Experimental results show that our proposed method can achieve the goal of emotion transfer while maintaining relatively high stability and timbre similarity. The future work will focus on extending the proposed method in fine-grained cross-speaker emotion transfer.
    \vfill\pagebreak
    
    \bibliographystyle{IEEEbib}
    \bibliography{strings,refs}
    
\end{document}